\documentclass{article}

\usepackage{PRIMEarxiv}

\usepackage[utf8]{inputenc} % allow utf-8 input
\usepackage[T1]{fontenc}    % use 8-bit T1 fonts
\usepackage{url}            % simple URL typesetting
\usepackage{microtype}      % microtypography
\usepackage{lipsum}
\usepackage{fancyhdr}       % header
\usepackage{graphicx}       % graphics
\graphicspath{{media/}}     % organize your images and other figures under media/ folder
\usepackage{subcaption}
\usepackage{hyperref}       % hyperlinks
\usepackage{url}            % simple URL typesetting
\usepackage{booktabs}       % professional-quality tables
\usepackage{amsfonts}       % blackboard math symbols
\usepackage{amssymb}
\usepackage{amsmath}
\usepackage{algorithm2e}
\usepackage{wrapfig}
\usepackage{nicefrac}       % compact symbols for 1/2, etc.
\title{Policy Gradient-Driven Noise Mask}

\author{
  Mehmet Can Yavuz, Yang Yang \\
  Department of Radiology and Biomedical Imaging\\ University of California - San Francisco, USA.\\
  \texttt{\{mehmetcan.yavuz, yang.yang4\}@ucsf.edu}
}

\begin{document}
\maketitle

\begin{abstract}
Deep learning classifiers face significant challenges when dealing with heterogeneous multi-modal and multi-organ biomedical datasets. The low-level feature distinguishability limited to imaging-modality hinders the classifiers' ability to learn high-level semantic relationships, resulting in sub-optimal performance. To address this issue, image augmentation strategies are employed as regularization techniques. While additive noise input during network training is a well-established augmentation as regularization method, modern pipelines often favor more robust techniques such as dropout and weight decay. This preference stems from the observation that combining these established techniques with noise input can adversely affect model performance.

In this study, we propose a novel \textit{pretraining} pipeline that learns to generate conditional noise mask specifically tailored to improve performance on multi-modal and multi-organ datasets. As a reinforcement learning algorithm, our approach employs a dual-component system comprising a very light-weight policy network that learns to sample conditional noise using a differentiable beta distribution as well as a classifier network. The policy network is trained using the reinforce algorithm to generate image-specific noise masks that regularize the classifier during pretraining. A key aspect is that the policy network's role is limited to obtaining an \textit{intermediate (or heated) model} before fine-tuning. During inference, the policy network is omitted, allowing direct comparison between the baseline and noise-regularized models.

We conducted experiments and related analyses on RadImageNet datasets. Results demonstrate that fine-tuning the intermediate models consistently outperforms conventional training algorithms on both classification and generalization to unseen concept tasks.
\end{abstract}

% keywords can be removed

\section{Introduction}

Image classification is a fundamental task in computer vision that involves assigning labels or categories to images based on their visual content. Traditional approaches to image classification have relied on conventional supervised learning techniques, where the model is trained on a labeled dataset. However, reinforcement learning (RL) has emerged as a promising enhancement to the training process, enabling models to learn optimal classification policies through interaction with an environment\cite{wiering2011reinforcement}\cite{sarkar2023rl}. In particular, policy gradient methods in RL offer a powerful framework for directly optimizing classification performance\cite{almahamid2021reinforcement}.

RL-based approaches offer a promising solution to address the challenges posed by image classification datasets\cite{wiering2011reinforcement, fujitake2024rl}. By learning optimal policies for feature extraction and classification through interaction with the environment, RL algorithms can adapt to the variations in low-level features and capture the relevant high-level semantic relationships \cite{sarkar2023rl}. This enables the development of more robust and accurate image classification models.

\begin{figure*}[h!]
    \begin{center}
    \includegraphics[width=0.9\textwidth]{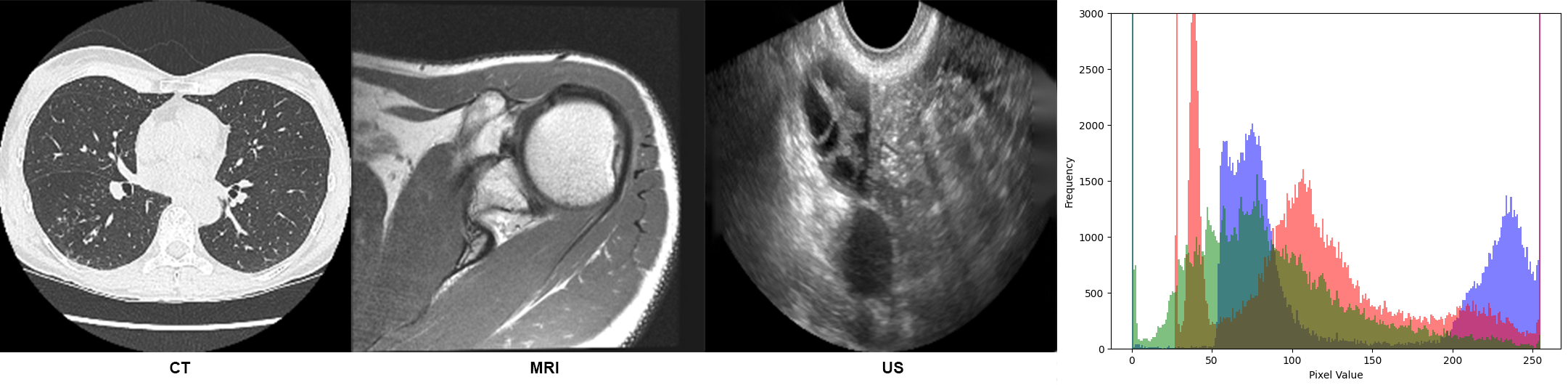}
    \end{center}
    \caption{Heterogeneity in medical imaging datasets (RadImageNet \cite{mei2022radimagenet}) across modalities and anatomical regions. (Left) CT scan of the lungs. (Second left) MRI scan of the shoulder. (Second Right) Ultrasound image of the ovary. (Right) Pixel intensity histograms as indicator for low-level image features of CT (blue), MRI (red), and ultrasound (green) images, illustrating variations in brightness, contrast, and noise levels.}
  \label{Fig:samples}
\end{figure*}

Medical imaging datasets, such as MRI scans, ultrasound images, and CT scans, often exhibit significant heterogeneity and variations in low-level features due to differences in acquisition protocols, imaging modalities, and anatomical regions an example illustration is in Figure \ref{Fig:samples}. These variations manifest as differences in brightness, contrast, and noise levels, which can be readily discerned even through visual inspection of image histograms. Such low-level distinguishability hinders the learning of high-level semantic relationships. The challenge is to capture and homogenize these low-level image features to improve classifier performance.

\vspace{-4mm}

Previously, injecting random noise into the input data during training has been proposed as a regularization approach to improve generalization \cite{bishop1995training}. However, the prevalence of other powerful regularizers like weight decay and dropout raises the question of whether additional noise-based regularization is beneficial \cite{koziarski2017image}. Moreover, prior work has highlighted the potential negative impact of noise-based regularization, an important consideration given that modern training pipelines typically omit additive or multiplicative noise \cite{akbiyik2023data}.

\begin{figure*}[h!]
    \begin{center}
    \includegraphics[width=0.9\textwidth]{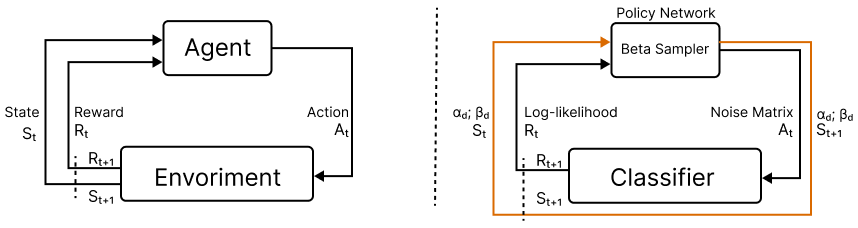}
    \end{center}
    \caption{Schematic diagram of Reinforcement learning. At left agent takes action and change the state in environment and gain reward. At right, beta sampler (policy network) generates noise matrix and classifier as a differentiable environment computes the log-likelihood and updates the state variables $\alpha$ and $\beta$.}
  \label{Fig:RL}
\end{figure*}

In this paper, we present several key contributions to tackle the challenges encountered by deep learning classifiers when working with heterogeneous multi-modal and multi-organ biomedical datasets. Our primary contribution is a novel \textit{pretraining} pipeline that learns to generate conditional noise masks specifically designed to enhance performance on these datasets. We propose a reinforcement learning system that consists of a lightweight policy network and a classifier network as shown in Figure \ref{Fig:RL}. During pretraining, the policy network is optimized to generate image-specific noise masks that regularize the classifier, effectively improving its performance on complex biomedical datasets. This approach enables the classifier to better handle the heterogeneity and multi-modality of the data, leading to more accurate and robust predictions in various biomedical applications.

\vspace{-4mm}

In addition to our novel pretraining pipeline, we perform comprehensive experiments and analyses on RadImageNet datasets to validate the effectiveness of our proposed approach. The results consistently show that fine-tuning the intermediate (or heated) models obtained through our \textit{pretraining} pipeline outperforms conventional training algorithms on both classification tasks and generalization to unseen concepts. This superior performance underscores the potential of our reinforcement learning-based noise regularization technique in enhancing the robustness and adaptability of deep learning classifiers when faced with challenging biomedical imaging scenarios.

In summary, our contributions significantly advance the state-of-the-art in deep learning for heterogeneous multi-modal and multi-organ biomedical datasets. By introducing a novel reinforcement learning-based approach, we effectively address the limitations of existing regularization techniques and provide a powerful tool for improving the performance of deep learning classifiers in complex biomedical imaging scenarios.

The remainder of this paper is organized as follows. The next sub-section provides an overview of related work in the field of medical image classification and reinforcement learning methodologies. Section 2 describes our proposed pretraining algorithm and background in detail, including training procedure. Section 3 presents the dataset, experimental setup and Section 4 includes results along with a discussion of the findings. Finally, Section 5 outlines and concludes the paper.

\textbf{Literature Review} Holmstrom et al. were among the pioneers in addressing this issue by using additive noise in back-propagation training, which can be seen as an early form of regularization to prevent overfitting \cite{holmstrom1992additive}. It is extended to the concept of the domain of speech recognition, demonstrating the effectiveness of noisy training for deep neural networks \cite{yin2015noisy}.

Bishop provided a theoretical foundation for training with noise, showing that it is equivalent to Tikhonov regularization, which adds a penalty term to the loss function to control the complexity of the model \cite{bishop1995training}. This concept is explored image recognition with deep neural networks in the presence of noise, demonstrating that distortions can be both a challenge and an opportunity for model training \cite{koziarski2017image}. Enhancing the generalisation abilities of neural networks (NNs) through integrating noise such as MixUp or Dropout during training has emerged as a powerful and adaptable technique. Despite the proven efficacy of noise in NN training, there is no consensus regarding which noise sources, types and placements yield maximal benefits in generalisation and confidence calibration. \cite{ferianc2024navigating}

It is proposed deep neural network architectures that are robust to adversarial examples, which are inputs crafted to deceive the model into making incorrect predictions \cite{gu2014towards}. This work is part of a broader effort to develop models that maintain high performance in the presence of input perturbations.

Dropout is introduced, a simple yet effective technique to prevent neural networks from overfitting \cite{srivastava2014dropout}. Dropout works by randomly omitting a subset of features during training, which encourages the model to learn more robust features. Further contributed to this field by introducing Cutout, a regularization method that randomly masks out sections of input images during training, forcing the network to focus on less prominent features \cite{devries2017improved}.

The use of semantic segmentation for masking and cropping input images has proven to be a significant aid in medical imaging classification tasks. The proposal of a novel joint-training deep reinforcement learning framework for image augmentation called Adversarial Policy Gradient Augmentation (APGA) that shows promising results on medical imaging tasks \cite{cheng2019adversarial}.

RadImageNet, represents a significant step forward in the domain of medical imaging \cite{mei2022radimagenet}. It is an open radiologic dataset designed to facilitate effective transfer learning in deep learning research. The MedMNIST Classification Decathlon, as presented by Yang et al., is a lightweight benchmark for medical image analysis designed to assess the capabilities of automated machine learning (AutoML) solutions \cite{yang2021medmnist}.

\section{Policy Gradient-Driven Noise Mask}

Our novel pretraining pipeline streamlines heterogeneous biomedical data through a three-pronged approach:
\begin{itemize}
    \item Adapting the noise masks to each specific image, accounting for variations in modality and organ type.
    \item Using reinforcement learning to optimize the noise masks for improved classifier performance.
    \item Separating the noise mask generation (pretraining) from the final model (fine-tuning), allowing for more flexible and effective regularization.
\end{itemize}

\begin{figure*}[th!]
  \begin{center}
    \includegraphics[width=\textwidth]{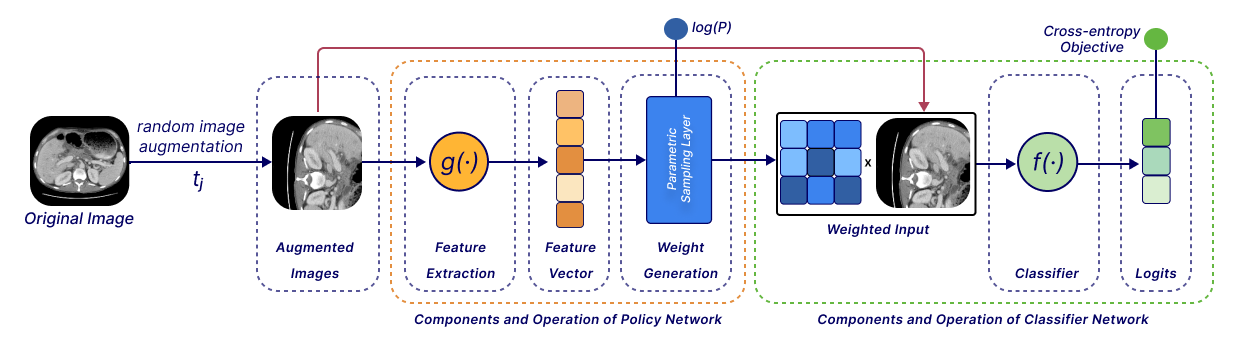}
  \end{center}
  \caption{Diagram of our proposed pipeline using deep learning, illustrating the process from original image through stochastic masking, feature extraction, beta sampling and classification to produce a prediction with a cross-entropy objective. The blue and green color parameters to compute objective function.}
  \label{Fig:rl_resnet}
\end{figure*}

We describe the mathematical formulation of the action taken by the policy network and the subsequent interaction with the environment, which leads to the computation of the loss function used for training. The order of steps follows the Figure \ref{Fig:rl_resnet} from left to right.

\textbf{Image preprocessing} is the initial stage, where raw images are prepared for further processing. This step may include normalization, resizing, and other image augmentation techniques. Mathematically, if $I$ represents the raw image, the preprocessing step can be represented as:
\begin{eqnarray}
    I_{pre} &=& t_{j}(I).
\end{eqnarray}
where $I_{pre}$ is the preprocessed image and the stochastic function $t_j(\cdot)$ is used to obtain random augmentations of the input image j.

\textbf{Action} Given an pre-processed input image $I_{pre}$, the policy network computes image specific parameters $\alpha_{image}$ and $\beta_{image}$ based on the dataset specific parameters represented by $\alpha_{dataset}$ and $\beta_{dataset}$. The updated parameters are obtained as follows:

\begin{equation}
\alpha_{image}, \beta_{image}, \alpha_{dataset}, \beta_{dataset} = \text{PolicyNet}(I, \alpha_{dataset}, \beta_{dataset}).
\end{equation}

\begin{figure*}[t!]
  \begin{center}
    \includegraphics[width=\textwidth]{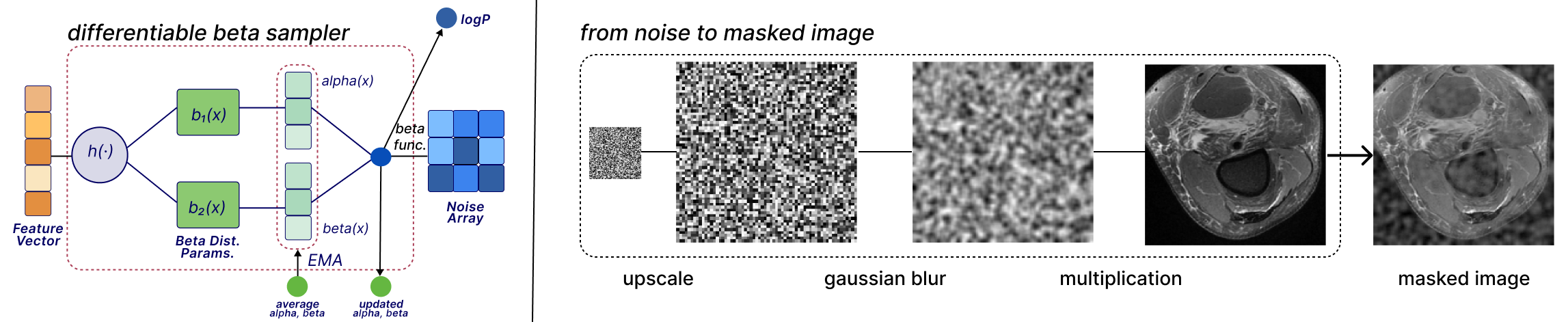}
  \end{center}
  \caption{Diagram of a policy network architecture (at left) showing the flow from input feature tensor to weighted tensor output. The network processes the input through a function $h(\cdot)$, projects the feature vector into Beta distribution parameters $b_1(x)$ and $b_2(x)$, derives alpha and beta values for the Beta function, calculates log probability (\text{log}P). The outputs are visualized over the dashed region as colorful circles. As an example (at right), the process of applying a stochastic masking to a medical image, showcasing the transformation from a noise matrix to the final masked image, which is part of the image processing pipeline involving steps such as noise matrix acquisition, upsampling, blurring, and applying. }
  \label{Fig:beta}
\end{figure*}

PolicyNet consists of feature extrator network $g(\cdot)$ and beta sampling operation based on features as shown in Figure \ref{Fig:rl_resnet}. The extracted feature vector let the beta sampler network to generate image specific noise mask as shown in Figure \ref{Fig:beta}.

These parameters are then exponentiated to ensure they are positive, as required by the Beta distribution:

\begin{equation}
\alpha' = e^{\alpha_{image}}, \quad \beta' = e^{\beta_{image}}.
\end{equation}

A Beta distribution $\mathcal{B}$ is then defined using these parameters and a mask $M$ is sampled from this distribution:

\begin{equation}
M \sim \mathcal{B}(\alpha', \beta').
\end{equation}

The mask is reshaped to match the dimensions of the input image and then interpolated if necessary to match the image dimensions.

We also introduce post-processing steps that involve upsampling and blurring as shown in Figure \ref{Fig:beta}. Firstly, we upsample the low-resolution noise matrix obtained from the policy network to obtain image regions with similar coefficient values, improving the correlation between neighboring pixels. Secondly, we apply a blurring operation to the upsampled noise matrix to avoid sharp transitions between different regions during the convolution operation, smoothing out the boundaries between areas with different noise values.

\textbf{Environment} The model takes the element-wise product of the input image $I$ and the mask $M_{\text{in}}$ to get the output:
\begin{equation}
O = \text{Model}(I_{pre} \odot M_{\text{in}}).
\end{equation}

The environment in this context is implicit. It consists of: (1) The distribution of input images $I$ that the model encounters. (2) The task-specific criteria that determine the reward, which is based on the model's output $O$ and the target labels $T$. (3) The Beta distribution from which the mask $M_{\text{in}}$ is sampled, which forms part of the action space of the policy.

\textbf{Policy Gradient Objective}
The policy gradient method is a fundamental approach in reinforcement learning for optimizing a policy function $\pi_\theta(a|s)$, where $\theta$ represents the policy parameters, $a$ denotes the action taken, and $s$ is the current state. The primary objective is to maximize the expected return $J(\theta)$. This is achieved by computing the gradient of the objective function with respect to the policy parameters $\theta$, given by:

\begin{equation}
\nabla_\theta J(\theta) = \mathbb{E}_{\tau \sim \pi\theta} \left[ \sum_{\eta=0}^{T} \nabla_\theta \log \pi_\theta(a_{\eta}|s_{\eta}) R(\tau) \right].
\end{equation}

Here, $\eta$ represents the time-step, and $\tau$ denotes a trajectory sampled from the policy $\pi_\theta$. This formulation allows for the update of policy parameters in the direction that increases the expected return. The policy gradient objective comprises two key components: (1) The log-probability output from the stochastic policy network. (2) The value of the reward function. For the specific application described, the log-probability of the occurrence of the mask $M_{\text{in}}$ under the Beta distribution is computed as:
\begin{equation}
\log(P) = \mathcal{B}.logP(M_{\text{in}}).
\end{equation}
This log-probability represents the output of the policy network. The reward function, in this case, is defined as the cross-entropy between the output logits and the corresponding labels for the input image (which represents the state). The loss function, combining these elements, is formulated as the mean of the product of the exponentiated log probabilities and the criterion applied to the output and the target:

\begin{equation}
\mathcal{L} = \text{mean}\left(\log\left(\sum e^{\log(P)}\right) \times \text{Criterion}(O, T)\right).
\end{equation}

where $O$ represents the output logits and $T$ denotes the target labels.
This formulation of the policy gradient method provides a framework for optimizing the policy in a reinforcement learning context, specifically tailored to the task of mask generation under a Beta distribution.

\textbf{Image and Dataset Specific Shape Parameters} The beta distribution is a continuous probability distribution defined on the interval $(0, 1)$ and is parameterized by two positive shape parameters, typically denoted as $\alpha$ and $\beta$. The probability density function (PDF) of the beta distribution with parameters $\alpha$ and $\beta$ is given by:

\begin{equation}
f(x; \alpha, \beta) = \frac{\Gamma(\alpha + \beta)}{\Gamma(\alpha)\Gamma(\beta)} x^{\alpha - 1} (1 - x)^{\beta - 1}, \quad 0 < x < 1.
\end{equation}

\noindent where $\Gamma(\cdot)$ is the gamma function, which is a generalization of the factorial function to complex numbers.

The beta distribution is a versatile distribution that can take on various shapes depending on the values of the shape parameters $\alpha$ and $\beta$. The extracted feature vectors are transformed by a projection layer h(·) (as shown in Figure \ref{Fig:beta}), which generates two separate tensors - alpha and beta - the beta distribution parameters. The beta distribution parameters - alpha and beta - are used to sample a random tensor via the beta function. This beta matrix contains values between 0 and 1. The logP output provides the log probability of this tensor under the beta distribution, which could be useful for training or interpreting the model. 

The alpha and beta parameters from the last iteration step are the input for the policy network and the update function for alpha and beta are determined by recursive exponential moving averages formula:

\begin{eqnarray}
    \alpha_{new,image} &=& \tau_{i} \cdot \alpha_{dataset} + (1 - \tau_{i}) \cdot \alpha_{image}. \\ 
    \beta_{new,image} &=& \tau_{i} \cdot \beta_{dataset} + (1 - \tau_{i}) \cdot \beta_{image}. \\
    \alpha_{dataset} &=& \tau_{d} \cdot \alpha_{dataset} + (1 - \tau_{d}) \cdot \alpha_{mean,image}. \\ 
    \beta_{dataset} &=& \tau_{d} \cdot \beta_{dataset} + (1 - \tau_{d}) \cdot \beta_{mean,image}.  
\end{eqnarray}

\noindent The alpha and beta parameters in the given formulas, in a recursive manner, are used to update the Beta distribution parameters for an image-level as well as dataset-level.

The formulas use exponential moving averages to update the $\alpha$ and $\beta$ parameters for both the current image and the overall dataset:
\begin{itemize}
    \item $\alpha_{dataset}$ and $\beta_{dataset}$ are updated by taking a weighted average of their previous values and the mean of the image-level parameters across the dataset ($\alpha_{mean,image}$, $\beta_{mean,image}$).     
    \item $\alpha_{new,image}$ and $\beta_{new,image}$ are updated by taking a weighted average of the dataset-level parameters ($\alpha_{dataset}$, $\beta_{dataset}$) and the current image-level parameters ($\alpha_{image}$, $\beta_{image}$). The weight is controlled by $\tau_i$, typically set to 0.9.
\end{itemize}

\begin{center}
\begin{algorithm}[h!]
\SetAlgoLined
\KwResult{Optimized model parameters}
 Randomly initialization model parameters and dataset level $\alpha_d$, $\beta_d$\;
 \While{not converged}{
    \textit{compute the $\alpha, \beta$\;}
    $\alpha_i, \beta_i, \alpha_d, \beta_d \leftarrow \text{update mask params}(image, \alpha_i, \beta_i, \alpha_d, \beta_d )$\;
    $\text{dist} \leftarrow \text{Beta}(\text{$\alpha_i, \beta_i$})$\;
    \textit{sample the noise matrix\;}
    $\text{noise matrix} \leftarrow \text{dist.sample()}$\;
    \textit{take action\;}
    $\text{cross entropy obj.} \leftarrow \text{classifier.step}(image, \text{noise matrix})$\;
    $\text{loss} \leftarrow \text{dist.log\_prob}(\text{noise matrix}) \times \text{cross entropy obj.}$\;
    $\text{loss.backward()}$\;
 }
 \caption{The policy gradient method algorithm for training policy network together with classifier mechanism.}
 \label{alg:beta}
\end{algorithm}
\end{center}

\textbf{The Algorithm} The pseudo-code for the proposed algorithm is given in Algorithm \ref{alg:beta}: \textbf{Initialization} The algorithm begins with the random initialization of model parameters. Additionally, dataset level parameters $\alpha_d$ and $\beta_d$ are also initialized. These parameters are crucial as they will be updated throughout the training process to optimize the regularization mechanism. \textbf{Optimization} The core of the algorithm is an iterative process that continues until a specified termination condition is met, which in this case is convergence. \textbf{Compute Alpha and Beta} At each iteration, the algorithm computes the values of $\alpha$ and $\beta$. These are the parameters of the Beta distribution, which is used to model the stochastic nature of the regularization mechanism. \textbf{Update Mask Parameters} The function `update mask params` is called with the current image and the alpha and beta parameters for both the individual and dataset levels. This function adjusts the parameters to better fit the data as the algorithm learns. \textbf{Sample Noise Mask} A noise mask is sampled from the Beta distribution parameterized by the updated $\alpha_i$ and $\beta_i$. This noise mask represents the probabilistic decisions made by the regularization mechanism at this stage of training. \textbf{Take Action} The classifier takes an action based on the current image and the sampled weight matrix. This step involves the classifier making a prediction which is then used to calculate the objective function, in this case, the cross-entropy loss. \textbf{Calculate Loss and Backpropagate} The loss is calculated by taking the log probability of the sampled weight matrix from the Beta distribution and scaling it by the cross-entropy objective. This loss is then backpropagated through the network to update the model parameters in a direction that minimizes the loss. The loop continues until convergence.

\section{Experiments}

\textbf{Datasets} For pre-training our models, we are utilizing stratified (train/val/test) split of RadImageNet \cite{mei2022radimagenet}, a large-scale multi-modal and multi-organ medical imaging dataset (see Appendix \ref{appendix:radimagenet}). The split we prepare let us to justify model performance using different training techniques. This diverse dataset should help our model learn general features and representations as well as dataset specific comparison. To evaluate the performance of our pre-trained model, we are using the enhanced MedMNIST Classification Decathlon \cite{yang2021medmnist}, which includes the original 10 MedMNIST datasets, as well as 2 additional MRI datasets and 1 ultrasound (US) dataset (seee Appendix \ref{appendix:medmnist}). This comprehensive benchmark covers a wide range of medical imaging tasks, modalities (e.g., X-ray, CT, MRI, US, Microscope, OCT), and anatomical regions. By assessing our model's performance on the enhanced MedMNIST Decathlon, we can determine how well it generalizes across various medical imaging applications.

\textbf{Implementation Details\footnote{\url{https://github.com/convergedmachine/Policy-Gradient-Driven-Noise-Mask}}\footnote{\url{https://huggingface.co/convergedmachine/Policy-Gradient-Noise-Mask}}} The training pipeline is configured through a set of hyper-parameters. The main model to train is Resnet-50 and the policy network is always a leight-weight network such as Resnet-10t. The batch size is 32 for each 8xV100 GPU with effective batch size is 256. Each training phase takes 90 epochs with SGD optimizer with learning rate 0.1, momentum 0.9 and weight decay 1e-4. The step learning rate scheduler reduce by 1/10 in 30 epochs cycle. The Resnet-10t policy network starts with 0.01 learning rate and same momentum and weight decay and using cosine annealing learning rate scheduler. The initial values for $\alpha_{dataset}$ and $\beta_{dataset}$ are random and sampled from normal distribution.

For unseen concepts, we extract the features from the freezed backbone network and using MLP for unseen concept generalization or using Logistic Regression for low-shot adaptability. Both pre-training phases use AdamW with default hyper-parameters until convergences (with no accuracy increment for 5 epochs).

\textbf{Baselines, Ablation Study and Optimal Model} For the ablation study, we determine the optimal hyperparameters for the upscale coefficient, kernel size, and stride through a systematic search. In ablation study, we employ Resnet-10t as the backbone and the policy network due to its compact size and ease of optimization. These experiments provide a comprehensive understanding of the model's behavior under different settings and help identify the most suitable configuration for the given task. In the evaluation task, we use Resnet50 as backbone and Resnet-10t as policy network models. Two model are compared: a baseline model and one improved training with a Gradient Policy technique. The performance (macro) metrics considered included \textit{Precision}, \textit{Recall}, \textit{F1 Score}, \textit{AUROC}, and \textit{Balanced Accuracy}.

\textbf{Case Analysis} We investigate the scenario where no upscaling is applied, and instead, pixel-level noise is directly introduced to the input.  \textit{Baseline Performance:} Without any noise model applied, the performance metrics serve as a baseline. \textit{Different Noise Models:} The application of Gaussian and Uniform noise models, following fine-tuning, at 32x32 and 64x64 noise matrix. \textit{Pure Noise Conditions:} Under conditions simulating pure noise (noise matrix equal to image size, 224x224).

\textbf{Generalization to unseen concepts} We evaluate the generalization performance of our model on unseen concepts using the protocol proposed \cite{sariyildiz2021conceptgeneralization}. For RadImageNet, except for the modalities, the samples and classes from MedMNIST are unseen concepts. The model is pretrained on three datasets: ImageNet-1K (IN1K), RadImageNet (RadIN), and RadImageNet using Gradient Policy (Grad. P. RadIN). We then extract features for each downstream dataset and evaluate the performance using a randomly initialized multi-layer perceptron.

\textbf{How fast can models adapt to unseen concepts?} We evaluate the model performance for unseen concepts using low-shots proposed in \cite{sariyildiz2021conceptgeneralization}. We use CT, MRI, US and XR samples from MedMNIST dataset and the sample numbers are 8, 16, 32, 64, 128 and 256, respectively. The model is pretrained on three datasets: ImageNet-1K (IN1K), RadImageNet (RadIN), and RadImageNet using Gradient Policy (Grad. P. RadIN).

\section{Results \& Discussion}

\begin{table}[t!]
\centering
\caption{Entropy and performance comparison of normal and heated models with different input types during policy gradient-driven training on RadImageNet. Bolds indicate the best balanced accuracy (higher is better). The underlined score is the best entropy (lower is better.)}
\label{tab:resnet_performance}
\small
\begin{tabular}{@{}lccccccc@{}}
\textbf{Model} & \textbf{Input} & \textbf{Entropy} & \textbf{Precision} & \textbf{Recall} & \textbf{F1} & \textbf{ROC} & \textbf{B.Acc.} \\
\midrule
(Normal) & Normalized & 0.3802 & 0.5929 & 0.5014 & 0.5226 & 0.9884 & 0.5014 \\ 
(CutMix) & Normalized & 0.3259 & 0.6022 & 0.5225 & 0.5444 & 0.9875 & \textbf{0.5225} \\ 
(Heated) & Normalized & 0.7156 & 0.2906 & 0.2825 & 0.2417 & 0.9497 & 0.2825 \\ 
(Heated) & Noisy & 0.3294 & 0.5967 & 0.5136 & 0.5402 & 0.9898 & 0.5136 \\ 
(Finetuned Heated) & Normalized & \underline{0.3177} & 0.6034 & 0.5211 & 0.5468 & 0.9900 & \textbf{0.5211} \\
\bottomrule
\end{tabular}
\label{table:heated_model}
\end{table}

We start by our experiment by explaining the intermediate (or heated) model concept experimentally in Table \ref{table:heated_model}. The results demonstrate the evolution of model entropy through different stages of training and input types. Initially, the heated model shows high entropy (0.7156) with normalized input, indicating a state of uncertainty. When trained with noisy input, the entropy decreases significantly (0.3294), suggesting improved robustness. After finetuning on the target domain, the heated model achieves the lowest entropy (0.3177) among all models, including normal and CutMix variants. This low entropy, combined with competitive performance metrics, indicates that the finetuned heated model has learned more effectively from the data compared to other approaches. Notably, while the finetuned heated model and CutMix model have similar accuracy (balanced accuracy of 0.5211 and 0.5225, respectively), the finetuned model's lower entropy suggests more confident and potentially more reliable predictions.

Table \ref{tab:resnet_performance2} compares the performance of lightweight Resnet-10 and baseline Resnet-50 models with and without the gradient policy technique in different augmentations settings (hard augmentation: CutMix \cite{yun2019cutmix}: soft augmentation: AugMix \cite{hendrycks2019augmix}, AutoAug \cite{cubuk2018autoaugment}, RandAug \cite{cubuk2020randaugment}). The intermediate model obtained by policy gradient technique is fine-tuned RadImageNet. For both model sizes, applying the gradient policy improves all metrics, compare to normal training and provides competitive results with CutMix.

\begin{table}[htb]
\centering
\caption{Performance Metrics for ResNet Models with Best Parameter Settings in RadImageNet}
\label{tab:resnet_performance2}
\small
\begin{tabular}{@{}lccccc@{}}
\toprule
\textbf{Technique} & \textbf{Precision} & \textbf{Recall} & \textbf{F1} & \textbf{ROC} & \textbf{B.Acc.} \\
\midrule
\multicolumn{6}{@{}r}{\textit{Baseline Model (ResNet-50)}} \\
\multicolumn{6}{@{}l}{\textit{Heavy Augmentations}} \\\midrule
Grad.P.$\bigodot$AutoAug $\to FT$ & 0.6034 & 0.5211 & 0.5468 & 0.9900 & \textbf{0.5211} \\
CutMix$\bigodot$AutoAug & 0.6022 & 0.5225 & 0.5444 & 0.9875 & \textbf{0.5225} \\
\multicolumn{6}{@{}l}{\textit{Augmentations}} \\\midrule
RandAug & 0.5877 & 0.5102 & 0.5313 & 0.9898 & 0.5102 \\
AugMix & 0.6037 & 0.5078 & 0.5326 & 0.9892 & 0.5078 \\
AutoAug & 0.5929 & 0.5014 & 0.5226 & 0.9884 & 0.5014 \\
\\
\multicolumn{6}{@{}r}{\textit{Lightweight Model (ResNet-10t)}} \\
\multicolumn{6}{@{}l}{\textit{Heavy Augmentations}} \\\midrule
Grad.P.$\bigodot$AutoAug $\to FT$ & 0.5672 & 0.4573 & 0.4838 & 0.9890 & \textbf{0.4573} \\
CutMix$\bigodot$AutoAug & 0.5405 & 0.4292 & 0.4523 & 0.9871 & 0.4292 \\
\multicolumn{6}{@{}l}{\textit{Augmentations}} \\\midrule
RandAug & 0.5110 & 0.4060 & 0.4275 & 0.9864 & 0.4060 \\
AugMix & 0.5111 & 0.4096 & 0.4297 & 0.9867 & 0.4096 \\
AutoAug & 0.5386 & 0.4262 & 0.4479 & 0.9871 & 0.4262 \\
\bottomrule
\end{tabular}
\\
\smallskip
\footnotesize\textit{Note:} Grad.P.: Gradient-Policy Noise Mask (this work) using best parameter setting: Noise Matrix 64x64, K:13 and S:6 Gaussian Blurring; FT: Finetuning; CutMix \cite{yun2019cutmix}, AugMix \cite{hendrycks2019augmix}, AutoAug \cite{cubuk2018autoaugment}, RandAug \cite{cubuk2020randaugment}. 
\ref{tab:resnet_performance2}
\end{table}

Table \ref{tab:unseen} shows the results on enhanced medical imaging datasets. It reports the F1 scores of feature extractor Resnet-50 backbone (pretrained with ImageNet-1K, RadImageNet, and RadImageNet intermediate model with gradient policy, respectively) and evaluated using MLP over extracted feature vectors. The intermediate model with gradient policy achieves the highest F1 scores on 11 out of 13 datasets, demonstrating strong generalization to unseen medical imaging concepts. The improvements are especially significant for the small-scale datasets. It is categorized into three: the small datasets is less than 10,000 samples,  mid-scale dataset range is between 10,000 and 30,000 samples, and the large-scale datasets are over 100,000 samples.

\begin{table}[thb]
\centering
\caption{The F1 scores (higher is better) indicate the generalization to unseen concepts by MedMNIST dataset \cite{yang2021medmnist} using Resnet-50 backbone. The results columns are for ImageNet, RadImageNet pretrained weights as well as Gradient Policy pretrained intermediate (heated) models. The t-distribution \%95 confidence scores are also provided for small (7 trials) and mid-scale (3 trials) datasets.}
\label{tab:unseen}
\begin{tabular}{lccc}
\hline
\textbf{Dataset}  & \textbf{IN1K}  & \textbf{RadIN}  & \textbf{Grad. P. RadIN} \\ \hline
\multicolumn{4}{c}{\textbf{Resnet-50}} \\ \hline
\multicolumn{4}{c}{\textit{Small Sets (7 Trials Averaged)}} \\ \hline
Breast US & 0.6662$\pm$0.0394 & 0.6866$\pm$0.0312 & \textbf{0.7649$\pm$0.0154} \\
Breast Cancer US & 0.5932$\pm$0.0137 & 0.5659$\pm$0.0170 & \textbf{0.6563$\pm$0.0111} \\
BrainTumor MR & 0.8431$\pm$0.0093 & 0.8275$\pm$0.0079 & \textbf{0.8880$\pm$0.0041} \\
Brain MR & 0.2258$\pm$0.0187 & 0.2257$\pm$0.0154 & \textbf{0.4351$\pm$0.0345} \\
Pneumonia XR & 0.8344$\pm$0.0093  & 0.8343$\pm$0.0085  & \textbf{0.8554$\pm$0.0101} \\ \hline
\multicolumn{4}{c}{\textit{Mid-scale Sets (3 Trials Averaged)}} \\ \hline
Blood Cell Mic.  & 0.9099$\pm$0.0024  & 0.8686$\pm$0.0066  & 0.\textbf{9261$\pm$0.0024} \\
Dermatoscope  & \textbf{0.4998$\pm$0.0179}  & 0.3596$\pm$0.0602  & 0.4643$\pm$0.0266 \\
OrganA CT  & 0.7272$\pm$0.0043  & 0.7339$\pm$0.0134  & \textbf{0.8108$\pm$0.0033} \\
OrganC CT  & 0.6579$\pm$0.0119  & 0.6772$\pm$0.0094  & \textbf{0.7438$\pm$0.0037} \\
OrganS CT  & 0.5622$\pm$0.0068  & 0.5649$\pm$0.0121  & \textbf{0.6170$\pm$0.0009} \\ \hline
\multicolumn{4}{c}{\textit{Large-Scale Sets (1 Trial)}} \\ \hline
Retinal OCT  & 0.5160  & 0.5220  & \textbf{0.5736} \\
Colon Pathology  & \textbf{0.8138}  & 0.7880  & 0.8042 \\
Tissue Mic.  & 0.3082  & 0.3447 & \textbf{0.4033} \\
\hline
\end{tabular}
\end{table}

\begin{figure*}[h!]
    \begin{center}
    \includegraphics[width=\textwidth]{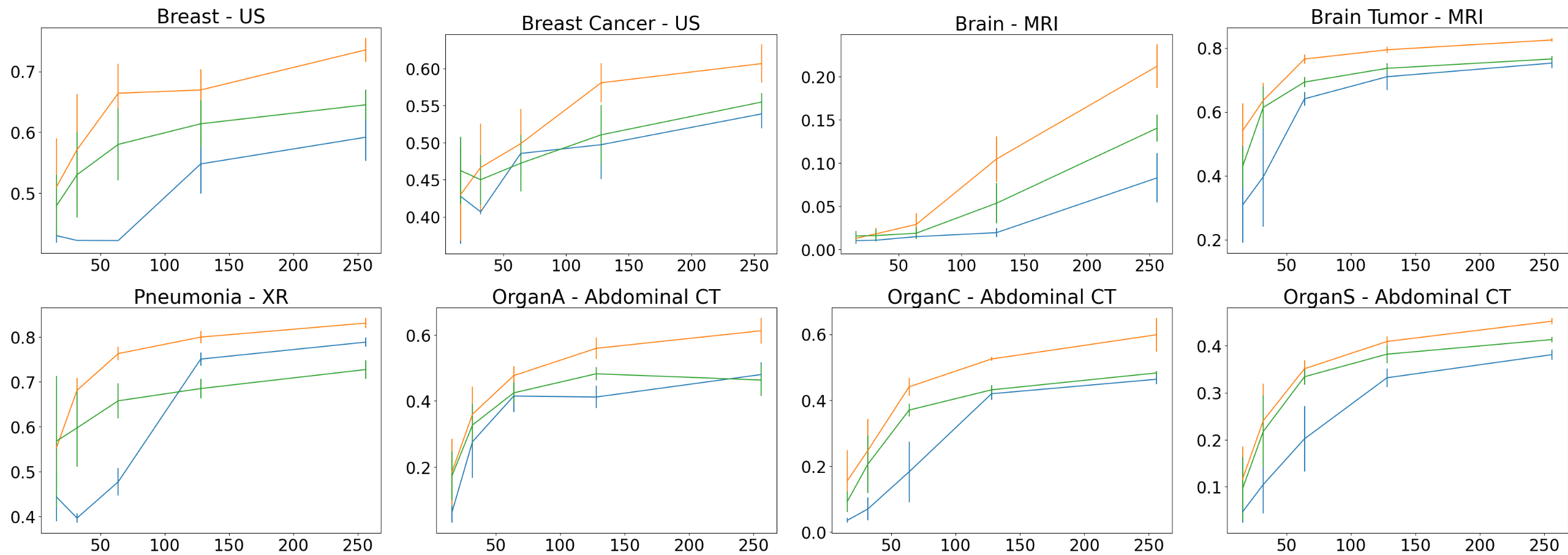}
    \end{center}  
    \caption{Each figure represents a different unseen but related dataset: Breast - US, Breast Cancer - US, Brain - MRI, Brain Tumor - MRI, Pneumonia - XR, OrganA, OrganC, OrganS - Abdominal CT. Different color curves corresponds to different pretrained model on few-shot adaptability. The orange curves represent the performance of Gradient Policy RadImageNet, the green curves show the results for ImageNet, and the blue curves indicate the performance of ImageNet pretrained models. The evaluation is carried out for 8, 16, 32, 64, 128, 256 samples and 10 trials in each sample size. The vertical lines are the t-statistics \%95 confidence interval.}
    \label{fig:unseen}
\end{figure*}

Figures \ref{fig:unseen} reports the few-shot adaptability of each CT, MRI and US modalities in 8 different dataset and 3 pre-trained networks (ImageNet-1K, RadImageNet, and RadImageNet with gradient policy) for (8, 16, 32, 64, 128, 256 samples and 10 trials in each sample size). The orange curves are pre-trained model with gradient policy which is consistently better in few-show adaptability.

Table \ref{tab:performance_metrics} shows the impact of different noise models and noise matrix sizes on the performance of the lightweight Resnet-10t model. Gaussian noise with a noise matrix of 64x64 yields the best precision, while uniform noise with a noise matrix of 64x64 gives the highest recall and accuracy. However, the differences between noise models are relatively small. Using pure noise matrix of size 224x224 leads to slightly lower but comparable performance to the baseline.

\begin{table}[thb]
\centering
\caption{Performance Metrics for Different Noise Models and Noise Matrix Sizes for RadImageNet \cite{mei2022radimagenet} by ResNet-10t model.}
\label{tab:performance_metrics}
\begin{tabular}{lcccccc}
\hline
\textbf{Features} & \textbf{Matrix} & \textbf{Precision} & \textbf{Recall} & \textbf{F1} & \textbf{AUROC} & \textbf{B. Accuracy} \\ \hline
\multicolumn{7}{c}{\textit{Different Noise Models}} \\ \hline
Baseline Model & - & 0.5386 & 0.4262 & 0.4479 & 0.9871 & 0.4262 \\
Gaussian + Blurring & 32x32 & 0.5354 & 0.4273 & 0.4494 & 0.9871 & 0.4273 \\
Uniform + Blurring  & 32x32 & 0.5206 & 0.4278 & 0.4480 & 0.9872 & 0.4278 \\
Gaussian + Blurring  & 64x64 & 0.5507 & 0.4279 & 0.4497 & 0.9874 & 0.4279 \\
Uniform + Blurring  & 64x64 & 0.5504 & 0.4299 & 0.4499 & 0.9861 & 0.4274 \\ \hline
\multicolumn{7}{c}{\textit{Pure Noise}} \\ \hline
Baseline Model & - & 0.5386 & 0.4262 & 0.4479 & 0.9871 & 0.4262 \\
Gaussian  & 224x224 & 0.5312 & 0.4236 & 0.4441 & 0.9864 & 0.4253 \\
Gaussian + Blurring  & 224x224 & 0.5396 & 0.4312 & 0.4534 & 0.9876 & 0.4312 \\
Pure Noise  & 224x224 & 0.5340 & 0.4314 & 0.4555 & 0.9876 & 0.4314 \\
Pure Noise + Blurring  & 224x224 & NaN & NaN & NaN & NaN & NaN \\ \hline
\end{tabular}
\end{table}

\begin{figure*}[h!]
  \begin{center}
    \includegraphics[width=\textwidth]{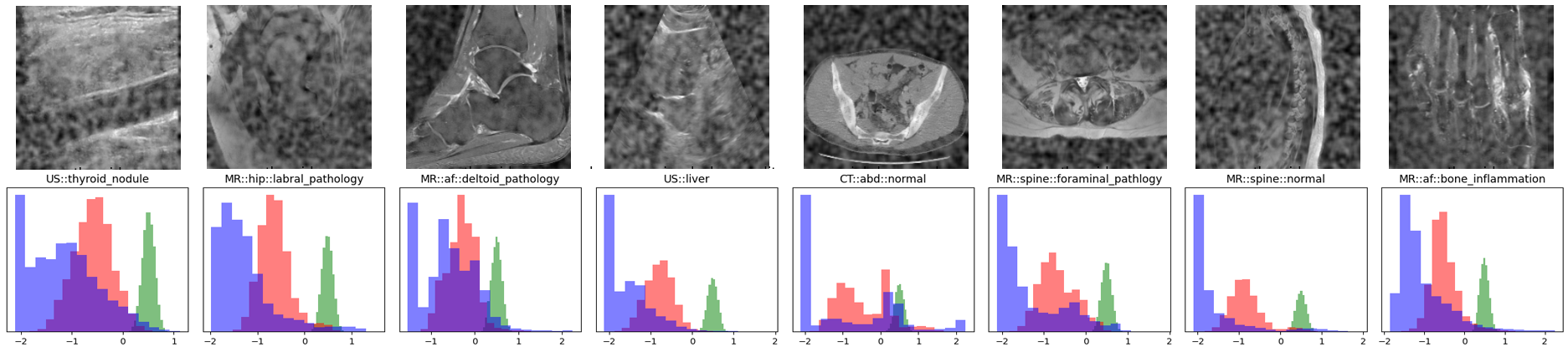}
  \end{center}
  \caption{Composite image displaying a series of medical scans from RadImageNet \cite{mei2022radimagenet} (1st row) with corresponding histograms (2nd row), where blue bars represent the original image pixel intensity distribution, red bars indicate the masked image pixel intensity distribution, and green bars show the noise pixel intensity distribution.}
  \label{fig:hist}
\end{figure*}

Figure \ref{fig:hist} presents medical scans with their respective histograms, indicating low-level features and pixel intensity distributions. The stochastic masking operation performed by the policy network modifies the skewness and center of distribution using pixel-wise multiplication, enhancing the image representation for the classifier and achieving a form of homogenization.

\begin{figure*}[h!]
  \begin{center}
    \includegraphics[width=\textwidth]{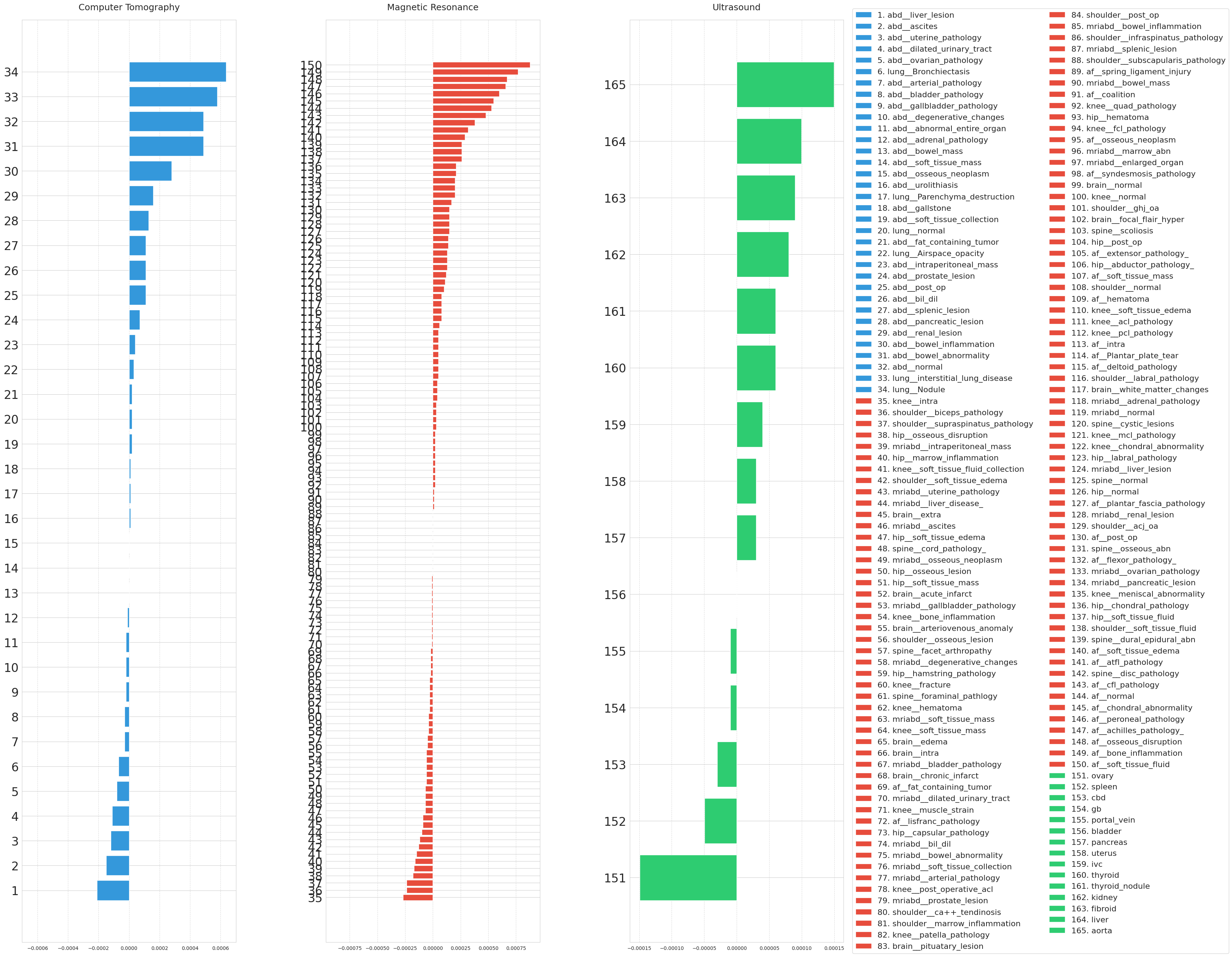}
  \end{center}
  \caption{Comparison of prediction differences between a Normal Model and a Gradient Policy Model across three medical imaging modalities: Computer Tomography (CT - blue), Magnetic Resonance Imaging (MRI - red), and Ultrasound (US - green). The horizontal bars represent the normalized difference in predictions for various anatomical structures and pathologies. Positive values indicate higher prediction rates by the Normal Model, while negative values show higher rates by the Gradient Policy Model. The red dashed line at zero serves as a reference point for equal performance between the two models.}
  \label{fig:acc}
\end{figure*}

Figure \ref{fig:acc} presents a comprehensive analysis comparing the gradient policy trained model to the conventional training approach. The diagram illustrates the performance tendency towards Policy Gradient per-labels

The results indicate that the proposed gradient policy technique consistently improves the performance of both lightweight and larger models for medical image classification. This suggests that the gradient policy helps the models learn more robust and generalizable features. However, the performance on some unrelated datasets, such as dermatoscope images and tissue microscopy, remains relatively low. Because the low-level image features are closer to natural images by ImageNet.

Interestingly, even though the noise matrix size has an effect on our proposed model, the known distributions such as Gaussian or Uniform does not effected. The pure noise condition does not substantially impact the performance metrics either. This could imply that the model is able to effectively handle different types of noise perturbations.

An important remark on the convergence is that the dataset level $\alpha$ and $\beta$ get `almost` uniform distribution shape after beta sampling operation almost always.

It is noteworthy that our experiments conducted on natural images sourced from the ImageNet-1K dataset did not yield superior accuracy compared to existing methods, despite our best efforts and the application of novel techniques.

\section{Conclusion}

In summary, the Gradient Policy technique has demonstrated its effectiveness in enhancing the performance and generalization capabilities of deep learning models in biomedical image analysis. The ablation study highlights the superiority of the our proposed training schema using Gradient Policy technique over the conventional training across all performance metrics. This is achieved by fine-tuning hyper-parameters such as grid size and Gaussian blurring parameters. Moreover, the technique's ability to improve the performance of larger models like Resnet-50 further underscores its versatility and scalability. On the other hand, the case analysis reveals that while variations in noise models such as using normal or uniform distribution or pure noise noise condition lead to minor performance differences, no statistically significant improvement is observed. 

The model's generalization performance on unseen concepts, evaluated using the protocol proposed by Sariyildiz et al., demonstrates the consistent superiority of the model pretrained on RadImageNet using Gradient Policy over models pretrained on ImageNet-1K and RadImageNet across all downstream datasets. This finding emphasizes the technique's ability to enhance the model's capacity to adapt to novel concepts and domains.

Furthermore, the low-shot adaptation performance on unseen concepts showcases the remarkable ability of the model pretrained on RadImageNet using Gradient Policy to quickly adapt to new concepts with limited samples, consistently outperforming models pretrained on ImageNet-1K and RadImageNet. This adaptability is crucial in the medical domain, where data scarcity and concept generalization are common challenges. The Gradient Policy technique not only improves the model's overall accuracy but also enables it to focus on relevant features and adapt quickly to unseen concepts with limited samples.

%Bibliography
\bibliographystyle{unsrt}  
\bibliography{references}  

\newpage
\appendix

\section{RadImageNet: Artifacts and The Refinement}
\label{appendix:radimagenet}

\textbf{RadImageNet} encompasses an extensive collection of biomedical images across three primary imaging modalities: CT, MRI and US. Image statistics are provided in Table 5. The dataset is reorganized and divided into three stratified splits: training, validation, and testing (Figure 8). Below is a summary based on the provided histogram and data distribution: \textit{Training Data}: 1,204,368 samples \textit{Validation Data}: 50,182 samples \textit{Test Data}: 100,364 samples. The dataset's split into train, validation, and test sets ensures robust model development and evaluation. The histograms highlight potential class imbalances that may need addressing through techniques like resampling or weighting during model training.

\begin{figure}[b]
    \centering
    \includegraphics[width=0.9\textwidth]{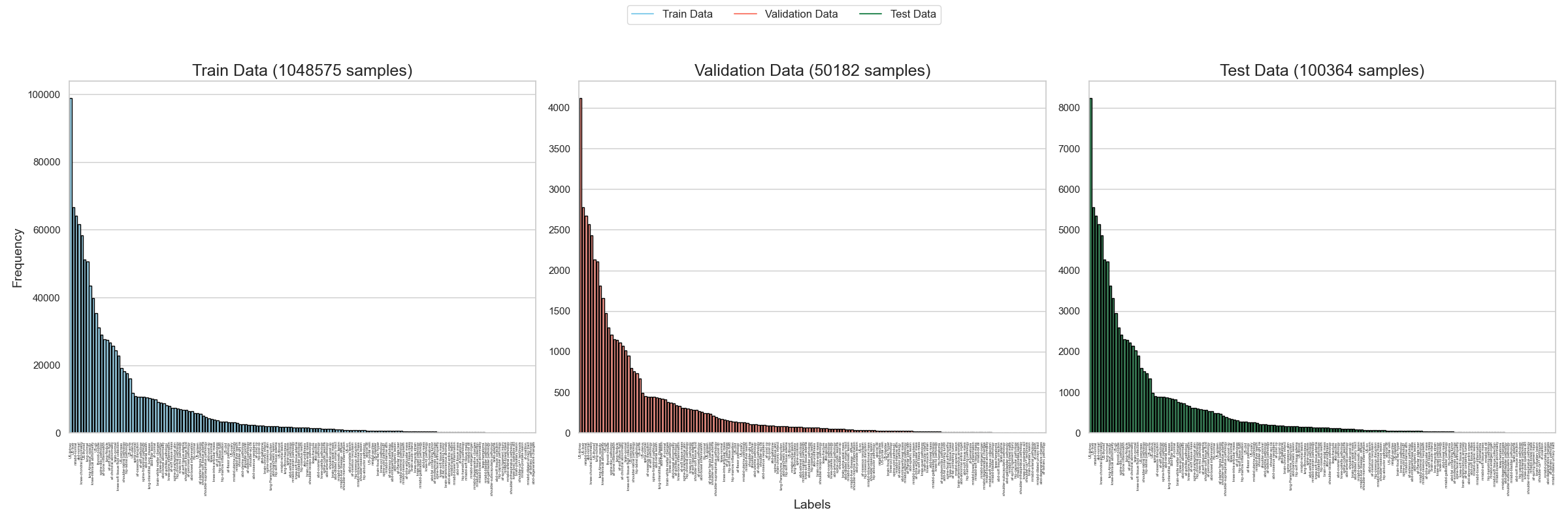}
    \caption{Histograms illustrating the frequency distribution of samples across categories for the Train (1,204,368 samples), Validation (50,182 samples), and Test (100,364 samples) datasets. The charts highlight potential class imbalances that may impact model training and evaluation.}
\end{figure}

\setlength\intextsep{1pt}
\begin{wraptable}{r}{0.55\textwidth}
    \centering
    \label{tab:image_statistics}
    \begin{tabular}{lccc}
        \toprule
        \textbf{Anatomy} & \textbf{\# Classes} & \textbf{Avg. Class} & \textbf{Total \#} \\
        \midrule
        \multicolumn{4}{l}{\textbf{A: CT (292,353 total images)}} \\
        Lung & 6 & 25,421 & 152,528 \\
        Abdomen/pelvis & 28 & 4,994 & 139,825 \\
        \midrule
        \multicolumn{4}{l}{\textbf{B: Ultrasound (389,885 total images)}} \\
        Thyroid & 2 & 46,300 & 92,599 \\
        Abdomen/pelvis & 13 & 22,868 & 297,286 \\
        \midrule
        \multicolumn{4}{l}{\textbf{C: MRI (672,675 total images)}} \\
        Knee & 18 & 9,975 & 179,555 \\
        Shoulder & 14 & 3,743 & 52,407 \\
        Spine & 9 & 7,964 & 71,674 \\
        Ankle/foot & 25 & 7,264 & 181,603 \\
        Abdomen/pelvis & 26 & 3,513 & 91,348 \\
        Brain & 10 & 4,467 & 44,671 \\
        Hip & 14 & 3,673 & 51,417 \\
        \bottomrule
    \end{tabular}
    \caption{Image statistics for CT, US, and MRI studies across various anatomical regions.}
\end{wraptable}

\subsection{Artifact Refinement Process}
The presence of artifacts (Figure 9) in ultrasound images necessitates a multi-step refinement process to enhance image quality:
\textit{Region of Interest (ROI) Isolation:} The algorithm processes specific regions within the images to identify and isolate areas affected by artifacts.
\textit{Thresholding with Otsu's Method:} Applied to create binary masks that distinguish foreground from background elements.
\textit{Morphological Operations:} Utilized to remove minor artifacts and noise, ensuring the preservation of essential image features.
\textit{Normalized Cross Correlation Coefficient (NCC):} Employed to assess similarity to known artifact templates, aiding in the identification of contaminated regions.
\textit{Adaptive Thresholding and Inpainting:} Dynamically identifies artifact-affected areas and employs the Telea algorithm for inpainting, effectively restoring image integrity.
\textit{Pixel-Level Outlier Detection:} Provides an additional layer of quality assurance by ensuring the final images are free from residual artifacts.

\begin{figure}[h]
  \begin{center}
    \includegraphics[width=\textwidth]{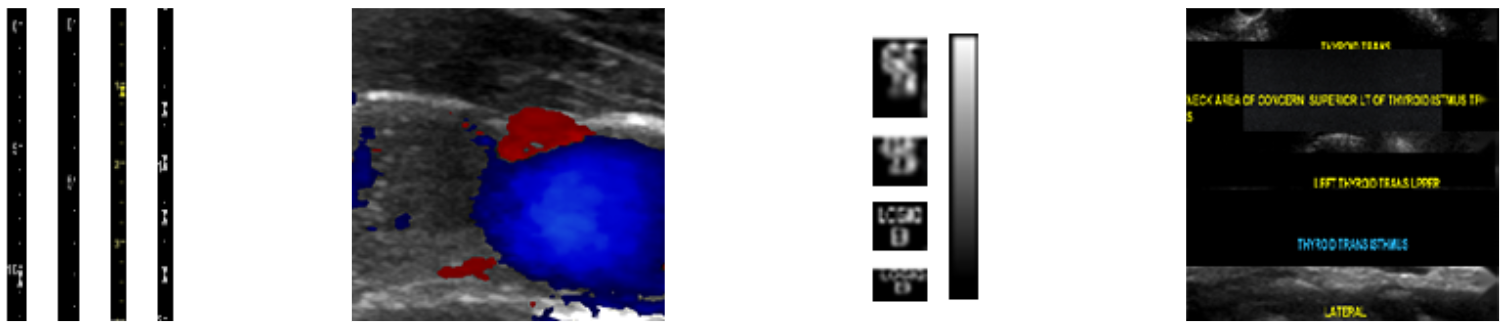}
  \end{center}
  \caption{Typical examples for the artifacts in ultrasound images: markers, segmentations, signs \& colorbar and texts,  respectively.}
  \label{fig:artifacts}
\end{figure}

\subsection{Automated Conversion Tool for RadImageNet}

In conjunction with this study, we introduce an automated tool designed to transform the existing RadImageNet dataset  \footnote{\url{https://github.com/convergedmachine/Refined-RadImagenet}.} 
\footnote{\url{https://huggingface.co/convergedmachine/RadImagenet}.} into its refined and stratified version without requiring manual intervention. This tool uses predefined masks to remove artifacts such as text annotations, markers, and segmentation masks that are commonly found in ultrasound images.

By automating the refinement process, this tool significantly reduces the time and effort required to prepare dataset for \textit{pretraining} purposes. It ensures that researchers can focus on developing fairly comparable diagnostic tools.

\section{Enhanced MedMNIST}
\label{appendix:medmnist}
MedMNIST is a comprehensive collection of standardized biomedical images designed for various analytical tasks in the medical field. This dataset has been expanded to include three new subsets, broadening the range of imaging modalities and classification challenges available to researchers. These additions complement the existing MedMNIST collections, offering a more diverse set of resources for developing and evaluating machine learning models across various medical imaging applications (Table 6). The enhanced MedMNIST collection, including both existing and new datasets, is accessible on Hugging Face \footnote{\url{https://huggingface.co/datasets/convergedmachine/Enhanced-MedMNIST}}

\begin{table}[htb!]
\small
\centering
\begin{tabular}{lllr}
\hline
\textbf{Name} & \textbf{Data Modality} & \textbf{Task (\#Cls)} & \textbf{\# Samples} \\ \hline
Brain Tumor Dataset & Magnetic Resonance & MC (3) & 3,064 \\
Brain Dataset & Magnetic Resonance & MC (23) & 1,600 \\
Breast Cancer & Ultrasound & BC (2) & 1,875 \\ \hline
\end{tabular}
\caption{The additional datasets for MedMNIST, the classification task including the number of classes (e.g., Multi-Class (MC) with 9 classes, Binary-Class (BC) with 2 classes), and the total number of samples contained within each dataset.}
\label{tab:dataset_summary}
\end{table}
\end{document}